\documentstyle[12pt]{article}

\def\beq{\begin{equation}}
\def\eeq{\end{equation}}
\def\bea{\begin{eqnarray}}
\def\eea{\end{eqnarray}}

\def\th{\theta}

\def\e{\epsilon}

\def\href#1#2{#2}

\begin{document}
\vspace*{-.6in}
\thispagestyle{empty}

\rightline{\small{\tt CALT-68-2364}}
\rightline{\small{\tt CITUSC/01-050}}

\rightline{\tt hep-th/0112238}

\vspace{.5in}

{\Large
\begin{center}
\vskip 2 cm

{\LARGE {A Note on Noncommutative D-Brane Actions}}

\vskip 2cm
{\large Calin Ciocarlie, Peter Lee, and Jongwon Park}

\vskip 1.2cm

\end{center}}

\begin{center}
\emph{ California Institute of Technology 452-48, Pasadena, CA  91125}

\vskip .7cm
{\tt calin, peter, jongwon@theory.caltech.edu}

\end{center}
\vspace*{1in}
\begin{center}
\textbf{Abstract}
\end{center}
\begin{quotation}
\noindent  We use the  nonabelian action of $N$ coincident D(-1)
branes in constant  background fields, in the $N \rightarrow
\infty$ limit, to construct noncommutative D-brane actions in an
arbitrary noncommutative description and comment on tachyon
condensation from this perspective.
\end{quotation}
\vfil
\newpage

\pagenumbering{arabic}

\section{Introduction and Review}

   It is a well known result that from the non-abelian Born-Infeld action of  infinitely many D(-1) instantons one can construct the background independent, $\Phi=-B$, description of noncommutative D-branes. Similarly, in \cite{Mukhi, Liu1, Liu2}, the same type of equivalence was shown for the Chern-Simons terms.
In \cite{IIKK}, it has been remarked that by placing D(-1)
instantons in a constant B-field one can construct noncommutative
D-branes with arbitrary noncommutativity. In this note, we clarify
this point by starting  from the action of $N$ coincident D(-1)
instantons in a constant $B$-field as given by \cite{Myers,
Tseytlin, Taylor}.  We show that such actions lead us to construct
D-brane actions in an arbitrary noncommutative description. The
map relating the Born-Infeld terms is seen to be consistent with
the map relating the Chern-Simons terms.  We use this result to
study tachyon condensation on a noncommutative D-brane with
arbitrary $\theta$.

We will now review some relevant results of \cite{Myers, Tseytlin}
and \cite{SeibergWitten}. For concreteness, we will assume
Euclidean space-time and maximal rank constant B-field along the
directions of a D$p$-brane.  We use the convention where
$2\pi\alpha^\prime=1$. Then the world-volume D$p$-brane action can
be described in noncommutative variables, i.e.
$[x^i,x^j]=i\th^{ij}$,  as

\beq
\hat S_{BI}=\frac{(2\pi)^{\frac{1-p}{2}}}{G_s} \int d^{p+1}x \sqrt{det(G+\hat{F}+\Phi)},
\eeq
where the * product is implicit in the above equation.
For abelian and constant $F$,  the Seiberg-Witten transformations relating $F$ to $\hat{F}$ are given by
\begin{equation}
F=\hat{F}\frac{1}{1-\theta \hat{F}}, \hspace{.5in} \hat{F}=\frac{1}{1+F\theta}F.
\end{equation}

For every closed string background characterized by the NS-NS
2-form $B$, the closed string metric $g$, and the closed string
coupling constant $g_s$, there is a continuum of descriptions
given by a choice of $\Phi$.  The open string metric $G$, the open
string coupling constant $G_s$ and the noncommutativity parameter
$\theta$ can be expressed in terms of closed string variables as
follows:
\bea \label{SW} && \frac{1}{G+\Phi}+\theta =  \frac{1}{g+B},  \\
\nonumber
 && G_s = g_s \left( \frac{det(G+\Phi)}{det(g+B)}\right)^{\frac{1}{2}}.
\eea

 Finally, let us review the main results of \cite{Myers, Tseytlin}.
The non-abelian Born-Infeld action describing $N$ (Euclidean)
coincident D$p$-branes in a closed string background defined by
$\phi, B'$ and $g$ is
\begin{equation}
S_{BI}= \frac{(2\pi)^{\frac{1-p}{2}}}{g_s}\int d^{p+1}\sigma Str\left(e^{-\phi}\sqrt{det(P[E_{ab}+E_{ai}(M^{-1}-\delta)^{ij}E_{jb}]+F_{ab})det(M^i_j)}\right),
\end{equation}
where $E \equiv g+B'$ and $\phi$ is the bulk dilaton. Furthermore,
${i,j}$ are indices for the transverse coordinates, ${a,b}$ are
indices for the coordinates parallel to the D-brane, and the $X$'s
are $N\times N$ matrices representing the transverse displacements
expressed in the static gauge. We also defined\footnote{Unlike in
~\cite{Myers}, we used the convention $F_{ab}=\partial_a A_b
-\partial_b A_a-i[A_a, A_b]$ in order to be consistent with the
definition of $\hat F$ in \cite{SeibergWitten}.} \beq M_j^i \equiv
\delta^i_j-i[X^i,X^k]E_{kj}. \eeq
 For  the non-abelian Chern-Simons action, we have
\begin{equation}
S_{CS}= \mu_p\int Str\left(P[e^{-i(i_{X}i_{X})}(\sum C^{(n)}e^{B'})]e^F\right),
\end{equation}
where $\mu_p$ is the RR charge of a D$p$-brane.  In the
aforementioned actions, the bulk fields should be considered
functionals of the $N\times N$ matrices $X$, and the trace should
be symmetrized between all expressions of the form $F_{ab},
D_aX^i, [X^i,X^j],$ and $X^k$. However, since we are only going to
consider D(-1) instantons in constant background  fields, these
details are irrelevant for our purposes.

More precisely, in the next two sections we consider an infinite number of D(-1) instantons with $\phi=0$ and where $g$ and $B'$ are constants. The presence of the $B'$ field will allow us to construct D-brane actions in an arbitrary noncommutative description.
  In section 2, we show that the Born-Infeld action of D(-1) instantons in a constant $B'$ field naturally leads to NC Born-Infeld action, where the $B$ field is identified as $B= B'+\th^{-1}$ for arbitrary noncommutativity  parameter $\th$.  Having shown this, the nonabelian generalization of the Chern-Simons action for an infinite number of D(-1) instantons should correspond to the NC Chern-Simons action in the same noncommutative description as the BI action.  This fact is confirmed in section 3.   In section 4, we comment on tachyon condensation using the connection to the matrix model. We find that in the presence of a  constant B-field, the vacuum becomes non-commutative.

\section{Born-Infeld Action}
In this section, we follow the line of thought in \cite{Seiberg}
and derive the equivalence of the nonabelian BI action of an
infinite number of D(-1) instantons and the BI action of a
noncommutative D$p$-brane in a general noncommutative description.
First consider the nonabelian BI action of $N$ D(-1) branes ($N
\rightarrow \infty$) in a constant $B'$-field:

\beq \label{MyersBI} S_{BI} = {2\pi \over g_s} Str \sqrt{
det_{ij}\left( {\delta_i}^j - i(g+{B'})_{ik} [X^k,X^j] \right) }.
\eeq We are interested in a particular classical configuration
given by \beq [x^i,x^j] = i \theta'^{ij}. \eeq The degrees of
freedom on the noncommutative D$p$-brane arise by expanding the
matrix variable $X^i$ around this classical configuration as
follows: \beq X^i = x^i + \theta'^{ij}\hat{A}'_j. \eeq Then, we
have \beq i[X^i,X^j] = (\theta'\hat{F}'\theta' - \theta')^{ij},
\eeq where \beq \hat{F}'_{ij}=
-i\th'^{-1}_{ik}[x^k,\hat{A}'_j]+i\th'^{-1}_{jk}[x^k,\hat{A}'_{i}]-i[\hat{A}'_{i},\hat{A}'_{j}].
\eeq We can reexpress $Tr$ over the Hilbert space as an integral
over the volume of noncommutative space by replacing \beq
 \label{TrInt}
 Tr \rightarrow {1\over (2\pi)^{(p+1)\over 2} Pf\th'} \int d^{p+1}x,
\eeq
where $Pf\th'$ is the Pfaffian. We write the action in terms of new variables,
\bea
S_{BI} &=&  {(2 \pi)^{1-p\over 2} \over g_s} \int \frac{d^{p+1}x}{Pf\theta'}  \sqrt{ det \left[ 1 - (g+B')(\theta'\hat{F}'\theta' - \theta')  \right] } \\
&=& {(2\pi)^{1-p\over 2} \over g_s} \int d^{p+1}x  \sqrt{ det\left[ \theta'^{-1} - (g+B')(\theta'\hat{F}' - {\bf 1})   \right] } \\
&=&  {(2\pi)^{1-p\over 2} \over g_s} \int d^{p+1}x  \sqrt{
det\left[ g+B'+\theta'^{-1} - (g+B')\theta'\hat{F}'   \right] }.
\eea We would like to compare this with the BI action of a
noncommutative D$p$-brane in a description with the same
noncommutativity parameter $\theta$ which appears in the above
action.  The NC BI action for a D$p$-brane is \beq \label{NCBI}
S_{NC BI} = {(2\pi)^{1-p\over 2} \over G_s} \int d^{p+1}x \sqrt{
det \left( G + \hat F +\Phi \right) }. \eeq Reexpressing it in
terms of closed string variables by using the relations (\ref{SW})
gives us \bea
S_{NC BI} &=& {(2\pi)^{1-p\over 2} \over g_s} \frac{\sqrt{det(g+B)}}{\sqrt{ det(G+\Phi)}} \int d^{p+1}x \sqrt{ det \left( G + \Phi + \hat{F} \right) } \\
&=& {(2\pi)^{1-p\over 2} \over g_s}  \int d^{p+1}x \sqrt{ det \left( g+B + (g+B)\frac{1}{G+\Phi} \hat{F} \right) } \\
&=& {(2\pi)^{1-p\over 2} \over g_s}  \int d^{p+1}x \sqrt{ det \left( g+B + (1-(g+B)\theta)\hat{F} \right) }.
\eea
We observe that (\ref{MyersBI}) agrees with (\ref{NCBI}) once we make
the following identifications:
\beq
 \label{iden1}
 \th=\th', \,\,\, \hat{F}=\hat{F}', \,\,\, B=B'+\th'^{-1}.
\eeq
  Notice that here  $\th$ is a free parameter, not fixed to be
$B^{-1}$ as in \cite{Seiberg}. By identifying $B'$ in the
nonabelian action for $N$ D(-1) instantons ($N \rightarrow
\infty$) with $B-\th^{-1}$, we can go to the noncommutative
description of D$p$-brane with arbitrary noncommutativity
parameter $\th$.  It is interesting to note that $\Phi$ takes the
following form in matrix-model-like variables: \beq
 \label{Phi}
 \Phi=-\th^{-1} \left(1+(g+B')^{-1}_A \th^{-1}\right),
\eeq
where $A$ denotes antisymmetrization.

\section{Chern-Simons Action}

If the nonabelian BI action for an infinite number of D(-1) instantons in a constant $B'$ field gives rise to the NC BI action with $B=B'+\th^{-1}$ and noncommutativity parameter $\th$, then we should expect the same identification relates the Chern-Simons term of the nonabelian action with that of the NC theory.
  This is precisely what occurs, and the Chern-Simons action for a D$p$-brane with a constant $B$ field and noncommutativity $\th$ can be expressed as the nonabelian CS action for an infinite number of D(-1) branes in a constant $B'$ field given by \cite{Myers}
\beq
 \label{MyersCS}
 S_{CS}={2\pi \over g_s} Str \left[e^{-i \left( {\rm
 i}_X {\rm i}_X \right)}
 \sum_n C^{(n)} e^{B'}  \right], \,\,\,\, B'=B-\th^{-1}.
\eeq Here ${\rm i}_X$ acts on an $n$-form $\omega^{(n)}$ as
 \beq
   {\rm i}_X \omega^{(n)} = {1\over (n-1)!} X^{\nu_1} \omega^{(n)}_{\nu_1
   \nu_2...\nu_n} dx^{\nu_2}... dx^{\nu_n}.
 \eeq
This provides a natural explanation of the rather surprising
result recently derived by \cite{Liu1}, where they express an
arbitrary NC CS action in terms of matrix-model like variables,
which turns out to be identical to (\ref{MyersCS}). For
simplicity, we follow the proof of \cite{Mukhi2} to show that the
nonabelian action gives rise to the NC action for D9-branes, where
we can ignore transverse scalar fields.  In that case, the NC CS
action is given by \cite{Liu1, Mukhi2} \beq
 \label{NCCS}
 S_{NCCS}=\mu_9 \int_x \sqrt{det(1-\th \hat F)}
      \sum_n C^{(n)} e^{B+\hat F (1-\th \hat F)^{-1}},
\eeq where $\mu_9 = (2\pi)^{-4}/g_s$ is the RR charge of a BPS
D9-brane. In terms of $Q=-\th+\th \hat F\th$, (\ref{NCCS}) can be
expressed as \beq
  S_{NCCS}=\mu_9 \int_x \sqrt{det(1-\th \hat F)}
      \sum_n C^{(n)} e^{B'}e^{-Q^{-1}}.
\eeq The nonabelian CS action for an infinite number of D(-1)
instantons ($\ref{MyersCS}$) naturally leads to the NC CS action
for D$p$-branes ($\ref{NCCS}$).   Expanding the action
($\ref{MyersCS}$) and using the fact that $i[X,X]=Q$ give terms of
the form \beq
  {2\pi\over g_s} Tr \left[  {(10-2r)!
  \over 2^{5-r} (s-r)! (5-r)! 2^{s-r} (10-2s)!} Q^{i_{2r+1} i_{2r+2}} ...
  Q^{i_{9} i_{10}} B'_{[i_{2r+1} i_{2r+2}} ... B'_{i_{2s-1} i_{2s}}
  C^{(10-2s)}_{i_{2s+1}...i_{10}]} \right],
\eeq where $[...]$ denotes antisymmetrization and $5 \ge s > r \ge
0$ . Employing the identity (\ref{TrInt}), one gets \bea
  \mu_9 \int d^{10} x {(10-2r)!
  \over 2^{5-r} (5-r)! (s-r)! 2^{s-r} (10-2s)! Pf\th} Q^{i_{2r+1} i_{2r+2}} ...
  Q^{i_{9} i_{10}} \times \\ \nonumber B'_{[i_{2r+1} i_{2r+2}} ... B'_{
  i_{2s-1} i_{2s}}
  C^{(10-2s)}_{i_{2s+1}...i_{10}]}.
\eea Finally, the above expression can be simplified to  \beq
  \label{CSterm}
  \mu_9 \int d^{10} x
  {PfQ (-1)^r \over Pf\th 2^s r! (s-r)! (10-2s)!} \e^{i_1 ... i_{10}} Q^{-1}_{i_1 i_2}
   ... Q^{-1}_{i_{2r-1} i_{2r}} B'_{[i_{2r+1} i_{2r+2}} ... B'_{i_{2s-1} i_{2s}}
  C^{(10-2s)}_{i_{2s+1}...i_{10}]}.
\eeq
One can immediately see that (\ref{CSterm}) are the terms coming from the
expansion of (\ref{NCCS}).  We have shown that our claim holds for the special
case $p=9$.  The general case  has been already considered in \cite{Liu1}.

Up to now, we have restricted the Ramond-Ramond fields to be
constants, but we can  generalize our procedure to the case where
the Ramond-Ramond fields are varying by writing the fields as
fourier transforms\footnote { See \cite{Yuji} for how to relate
the currents expressed in matrix model language to those in
noncommutative gauge theory.} such that \beq
 S_{CS}={2\pi \over g_s}\int d^{10}q Str \left[e^{-i \left( {\rm
 i}_X {\rm i}_X \right)}
 \sum_n C^{(n)}(q) e^{B'}e^{i q \cdot X}  \right],  \,\,\,\, B'=B-\th^{-1}.
\eeq To conclude, motivated by the identification relating the
nonabelian BI action of D(-1) instantons to the BI action of
D$p$-branes in the last section, we have proposed and verified
that the NC CS action of a D$p$-brane with arbitrary
noncommutativity and varying Ramond-Ramond fields can be derived
from considering the nonabelian CS action for an infinite number
of D(-1) branes after identifying $B'=B-\th^{-1}$.

\section{Tachyon Condensation}
In this section, we study tachyon condensation \cite{Sen, Sen2} in
open string theory via the matrix model connection as in
\cite{Seiberg, Strominger}.  Using the results of the previous
sections, we study this from an arbitrary noncommutative
description. The effective action on a single unstable
noncommutative D-brane is

\beq
  {(2\pi)^{1-p\over 2} \over G_s} \int d^{p+1}x \left[ V(T) \sqrt{ det \left( G + \hat F +\Phi \right)} +\sqrt{G} f(T) G^{ij} D_i T D_j T+...\right].
\eeq In terms of matrix-model-like variables, $\Phi$ is given by
(\ref{Phi}), while the open string metric is $G= -\th^{-1}
(g+B')^{-1}_S \th^{-1}$ where $S$ denotes symmetrization. Thus the
effective action can be written as
 \bea {2\pi \over g_s}  Tr \left[ V(T) \sqrt{ det_{ij}\left(
  {\delta_i}^j - i(g+{B'})_{ik} [X^k,X^j] \right)} \right. -\\ \nonumber
  \left. f(T)
  \sqrt{\frac{g}{g-B'} } (g-B' g^{-1} B')_{ij} [X^i,T][X^j,T]
  \right].
 \eea
Following \cite{Seiberg}, we assume that $V(T)$ has a unique
minimum at $T=T_c$ ($T_c$ proportional to the unit matrix). The
end-point of tachyon condensation obtained by minimizing the
Born-Infeld term\footnote{ One can write $\;
det\left(1-i(g+B')[X,X]\right)=det(g+B')det \left(
(g+B')^{-1}_S+(g+B')^{-1}_A-i[X,X]\right)$.} is characterized by
$X=X_c$ satisfying
 \beq
  \label{vac}
  [X_c,X_c]= -i (g+B')^{-1}_A.
 \eeq
We expect this  minimum to be exact,  in the sense that even if
there are corrections to the symmetrized-trace proposal for
non-abelian Born-Infeld action, these corrections are of the type
$[F,F]$ and $DF$, so they are irrelevant for our solution. For
$B'=0$, (\ref{vac}) implies that the vacuum is
commutative. In the presence of a non-zero  constant NS-NS field
it changes to a non-commutative state given by (\ref{vac}).

\section{Discussion}

It is interesting to analyze the Seiberg-Witten limit
in this context.  In this limit, using the conventions of
\cite{SeibergWitten}, we have
\bea
 && g \sim \epsilon, \\ \nonumber
 && \alpha^\prime \sim \epsilon^{1\over 2}, \\ \nonumber
 && B=B^{(0)} + \epsilon B^{(1)} + O(\epsilon^{2}), \\ \nonumber
 && \th^{-1} = B^{(0)}+O(\epsilon).
\eea
Keeping track of $\alpha^\prime$, we see that
$2\pi\alpha^\prime B'= 2\pi\alpha^\prime(B-\th^{-1})$ scales as
$O(\epsilon^{3/2})$ and $[X,X]\over 2\pi\alpha'$ goes like
$O(\epsilon^{-1/2})$. Hence, in the Seiberg-Witten limit, the
Born-Infeld action reduces to \beq \label{BISW}
 S_{BI} ={\pi \over g_s}\left[ i B'_{ij}Tr [X^i,X^j] - {1\over 2 (2 \pi \alpha')^2}
 g_{ik} g_{jl}  Tr \left([X^i,X^j] [X^k,X^l]\right)\right]+O(\epsilon^{1/ 2})+constant,
\eeq where the second term is the usual potential of the matrix
model. Furthermore, the nonabelian Chern-Simons action takes the
standard matrix model form
\beq
 S_{CS}={2\pi \over g_s} Str \left[e^{-{i\over 2\pi \alpha'} \left( {\rm
 i}_X {\rm i}_X \right)}  \sum_n C^{(n)}  \right],
\eeq where we assumed appropriate scaling of RR potentials,
$C^{(n)}$, such that the limit is well-defined.

Finally, let's remark that since $B=B'+\th^{-1}$, the freedom of
description of NC D$p$-branes translates in the matrix model like
variables into how one separates the $B$-field into the external
part $B'$ and the internal part $\th^{-1}$. The internal part,
$\th^{-1}$, is generated by the configuration of D(-1) instantons
and $B'$ corresponds to the external field imposed on them.

\begin{center}
\bf{Acknowledgments}
\end{center}
We have greatly benefitted from discussions with Iosif Bena, Iouri
Cheplev, Sangmin Lee, John Schwarz and especially Yuji Okawa.

\bibliography{action}

\begingroup\raggedright\begin{thebibliography}{10}

\bibitem{Mukhi}
S.~Mukhi and N.~V. Suryanarayana, ``Gauge-Invariant Couplings of Noncommutative
  Branes to Ramond-Ramond Backgrounds,'' {\em JHEP} {\bf 0105} (2001) 023,
  \href{http://xxx.lanl.gov/abs/hep-th/0104045}{{\tt hep-th/0104045}}.

\bibitem{Liu1}
H.~Liu and J.~Michelson, ``*-Trek III: The Search for Ramond-Ramond
  Couplings,'' {\em Nucl. Phys.} {\bf B614} (2001) 330--366,
  \href{http://xxx.lanl.gov/abs/hep-th/0104139}{{\tt hep-th/0104139}}.

\bibitem{Liu2}
H.~Liu and J.~Michelson, ``Ramond-Ramond Couplings of Noncommutative
  D-branes,'' {\em Phys. Lett.} {\bf B518} (2001) 143--152,
  \href{http://xxx.lanl.gov/abs/hep-th/0104139}{{\tt hep-th/0104139}}.

\bibitem{IIKK}
N.~Ishibashi, S.~Iso, H.~Kawai, and Y.~Kitazawa, ``Wilson Loops in
  Noncommutative Yang Mills,'' {\em Nucl. Phys.} {\bf B573} (2000) 573--593,
  \href{http://xxx.lanl.gov/abs/hep-th/9910004}{{\tt hep-th/9910004}}.

\bibitem{Myers}
R.~C. Myers, ``Dielectric-Branes,'' {\em JHEP} {\bf 9912} (1999) 022,
  \href{http://xxx.lanl.gov/abs/hep-th/9910053}{{\tt hep-th/9910053}}.

\bibitem{Tseytlin}
A.~A. Tseytlin, ``On non-abelian generalisation of Born-Infeld action in string
  theory,'' {\em Nucl. Phys.} {\bf B501} (1997) 41--52,
  \href{http://xxx.lanl.gov/abs/hep-th/9701125}{{\tt hep-th/9701125}}.

\bibitem{Taylor}
W.~Taylor and M.~V. Raamsdonk, ``Multiple Dp-branes in Weak Background
  Fields,'' {\em Nucl. Phys.} {\bf B573} (2000) 703--734,
  \href{http://xxx.lanl.gov/abs/hep-th/9910052}{{\tt hep-th/9910052}}.

\bibitem{SeibergWitten}
N.~Seiberg and E.~Witten, ``String Theory and Noncommutative Geometry,'' {\em
  JHEP} {\bf 9909} (1999) 032,
  \href{http://xxx.lanl.gov/abs/hep-th/9908142}{{\tt hep-th/9908142}}.

\bibitem{Seiberg}
N.~Seiberg, ``A Note on Background Independence in Noncommutative Gauge
  Theories, Matrix Model and Tachyon Condensation,'' {\em JHEP} {\bf 0009}
  (2000) 003, \href{http://xxx.lanl.gov/abs/hep-th/0008013}{{\tt
  hep-th/0008013}}.

\bibitem{Mukhi2}
S.~Mukhi and N.~V. Suryanarayana, ``Ramond-Ramond Couplings of Noncommutative
  Branes,'' \href{http://xxx.lanl.gov/abs/hep-th/0107087}{{\tt
  hep-th/0107087}}.

\bibitem{Yuji}
Y.~Okawa and H.~Ooguri, ``An Exact Solution to Seiberg-Witten Equation of
  Noncommutative Gauge Theory,'' {\em Phys. Rev} {\bf D64} (2001) 046009,
  \href{http://xxx.lanl.gov/abs/hep-th/0104036}{{\tt hep-th/0104036}}.

\bibitem{Sen}
A.~Sen, ``Descent Relations Among Bosonic D-branes,'' {\em Int. J. Mod. Phys.}
  {\bf A14} (1999) 4061--4078,
  \href{http://xxx.lanl.gov/abs/hep-th/9902105}{{\tt hep-th/9902105}}.

\bibitem{Sen2}
A.~Sen, ``Universality of the tachyon potential,'' {\em JHEP} {\bf 9912} (1999)
  027, \href{http://xxx.lanl.gov/abs/hep-th/9911116}{{\tt hep-th/9911116}}.

\bibitem{Strominger}
R.~Gopakumar, S.~Minwalla, and A.~Strominger, ``Symmetry Restoration and
  Tachyon Condensation in Open String Theory,'' {\em JHEP} {\bf 0104} (2001)
  018, \href{http://xxx.lanl.gov/abs/hep-th/0007226}{{\tt hep-th/0007226}}.

\end{thebibliography}\endgroup
\bibliographystyle{ssg}
\end{document}